\pdfoutput=1
\documentclass{article}
\usepackage{spconf,amsmath,graphicx,amssymb}
\usepackage{multirow}
\usepackage{threeparttable}
\usepackage{booktabs}
\usepackage{pifont}
\usepackage{url}
\usepackage{xcolor}

\usepackage{pifont}
\newcommand{\cmark}{\ding{51}}
\newcommand{\xmark}{\ding{55}}
\title{Advancing Multi-talker ASR Performance with Large Language Models}
\name{\shortstack{Mohan Shi$^{*}$, Zengrui Jin, Yaoxun Xu, Yong Xu$^{\dagger}$, Shi-Xiong Zhang, \\Kun Wei, Yiwen Shao, Chunlei Zhang, Dong Yu\thanks{*Done during internship at Tencent AI Lab. (shimohan@g.ucla.edu)}
\thanks{$^{\dagger}$Corresponding author. (lucayongxu@global.tencent.com)
}}}
\address{
    Tencent AI Lab, Bellevue, USA
    }
    
\begin{document}
%
\maketitle
\begin{abstract}
Recognizing overlapping speech from multiple speakers in conversational scenarios is one of the most challenging problem for automatic speech recognition (ASR). Serialized output training (SOT) is a classic method to address multi-talker ASR, with the idea of concatenating transcriptions from multiple speakers according to the emission times of their speech for training. However, SOT-style transcriptions, derived from concatenating {\color{black}multiple related utterances in a conversation}, depend significantly on modeling long contexts. 
{\color{black}{Therefore, compared to traditional methods that primarily emphasize encoder performance in attention-based encoder-decoder (AED) architectures, a novel approach utilizing large language models (LLMs) that leverages the capabilities of pre-trained decoders may be better suited for such complex and challenging scenarios.}}
In this paper, we propose an LLM-based SOT approach for multi-talker ASR, leveraging pre-trained speech encoder and LLM, fine-tuning them on multi-talker dataset using appropriate strategies. Experimental results demonstrate that our approach surpasses traditional AED-based methods on the simulated dataset LibriMix and achieves state-of-the-art performance on the evaluation set of the real-world dataset AMI, outperforming the AED model trained with 1000 times more supervised data in previous works.

\end{abstract}
\vspace{-0.1cm}

\section{Introduction}
\label{sec:intro}
\vspace{-0.2cm}
Although automatic speech recognition (ASR)~\cite{li2022recent,GulatiQCPZYHWZW20,yao2023zipformer} has achieved excellent performance in quiet, single-speaker scenarios, it still faces significant challenges in multi-talker conversational scenarios, especially in the case of overlapping speech. 
To overcome this challenge, a series of multi-talker ASR approaches have been proposed~\cite{chen2017progressive,yu2017recognizing,chang2019mimo,ZhangCQW20,kanda2020serialized}. One of the most representative methods is serialized output training (SOT)~\cite{kanda2020serialized,KandaYGWMCY21,ShiD0YLZ0023}. The core idea of SOT is to concatenate the transcriptions of multiple speakers in the order of their speech emission times, separated by a speaker change symbol. Compared to permutation invariant training (PIT)~\cite{yu2017recognizing,chang2019mimo,ZhangCQW20}, SOT avoids the limitation on the maximum number of speakers, models the dependencies in multi-talker content, and reduces computational complexity, resulting in better performance on multi-talker ASR task.

{\color{black}However, in SOT-style transcriptions, the concatenation of related content from multiple speakers, coupled with the relatively poor grammatical structure of sentences in meeting discussions, necessitates strong long-context awareness and cross utterance modeling. This is precisely what previous SOT methods based on attention-based encoder-decoder (AED)~\cite{kanda2020serialized}, which relied more on encoder performance, lacked, leading to performance bottlenecks.}
For instance, in~\cite{KandaYWGWMCY21}, despite using 900K hours of large-scale simulated data for pre-training, the word error rate on the AMI~\cite{CarlettaABFGHKKKKLLLMPRW05} meeting corpus still reached 21.2\%.

Large language models (LLMs)~\cite{abs-2211-05100,abs-2302-13971,abs-2307-09288,chiang2023vicuna}, trained on vast amounts of text data, possess unparalleled capabilities in understanding and generating natural language. Their proficiency in long-context awareness makes them exceptionally well-suited for SOT-style transcriptions. Therefore, the combination of LLM and SOT-based multi-talker ASR is a perfect match.
{\color{black}A series of LLM-based ASR works~\cite{WangHSWCCCZSRZYPSSW23,abs-2307-11795,abs-2310-13289,abs-2311-07919,abs-2402-08846,abs-2405-02132} have been conducted, which, in contrast to traditional AED methods that focus on encoder performance, tend to treat the speech foundation encoder~\cite{BaevskiZMA20,HsuBTLSM21,ChenWCWLCLKYXWZ22,RadfordKXBMS23} in LLM-based models as a tool for extracting embedding. The speech embedding then serve as prompt for the LLM, relying on the powerful decoder-only LLM to generate transcription.}
These studies have shown that this approach can match or slightly outperform traditional AED methods in simple single-speaker ASR tasks~\cite{abs-2307-11795,abs-2402-08846}. However, in these works, the performance advantage of the LLM-based methods is not particularly pronounced, indicating that LLM-based models, {\color{black}with their powerful decoders,} have not fully realized their potential in handling speech tasks in simple scenarios.

Therefore, in this paper, motivated by the potential of powerful LLMs to handle challenging speech tasks in complex scenarios and the natural compatibility of LLMs with SOT, we propose an LLM-based approach for multi-talker ASR. Similar to previous LLM-based ASR works, we employ a architecture comprising a pre-trained speech encoder, a projector, and an LLM. In previous works, various training strategies have been employed. For example, in~\cite{abs-2307-11795}, low-rank adaptation (LoRA)~\cite{HuSWALWWC22} was introduced into the LLM to facilitate efficient fine-tuning, and all three components were fine-tuned together in a single stage. In~\cite{abs-2402-08846}, LoRA was not introduced, and the encoder was frozen while training only the projector, which also yielded satisfactory results. In~\cite{abs-2405-02132}, a multi-stage fine-tuning approach was used to better align the modalities of speech and text. In this paper, we compared the aforementioned training strategies on the simulated LibriMix dataset and synthesized the best practices to propose the most suitable strategy, which made our LLM-based method surpass the AED-based approach. On the evaluation set of the real-world meeting corpus AMI, the proposed LLM-based method not only surpasses AED-based methods trained with the same amount of data but also remarkably outperforms the AED model trained on an enormous scale of 900K hours (1000 times more) of supervised data, achieving state-of-the-art. This astounding result demonstrates the immense potential of LLM-based models in handling speech processing tasks in challenging scenarios.

\section{Method}
\label{sec:method}
\vspace{-0.1cm}
\subsection{Serialized Output Training}
\begin{figure}[h]
 	\centering
 	\includegraphics[width=1\linewidth]{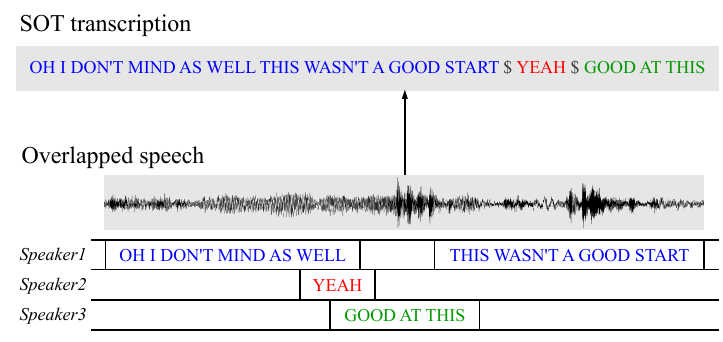}
 	\caption{
 	SOT transcription following speaker-wise FIFO
 	}
 	\label{fig:sot}
\end{figure}

\vspace{-0.1cm}
Serialized output training (SOT) is an elegant method to address multi-talker ASR. During the training stage, the transcriptions of different speakers are concatenated using a speaker change symbol to create the reference transcription for the overlapping speech. The concatenation order follows the emission time of each speaker, known as first-in first-out (FIFO). For example, as shown in Fig.~\ref{fig:sot}, in the case of three speakers, the reference transcription $Y$ is given as $R=\{r_1^1, \cdots, r_{N_1}^1, \$ , r_1^2, \cdots, r_{N_2}^2, \$ , r_1^3, \cdots, r_{N_3}^3 \}$, where $r_i^j$ represents the $i$-th token of the $j$-th speaker, $N_j$ represents the number of tokens in the $j$-th speaker, and ``$\$$'' represents the speaker change symbol.

\begin{figure}[t]
 	\centering
 	\includegraphics[width=0.75\linewidth]{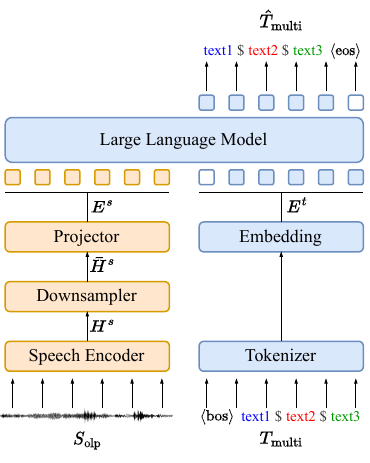}
 	\vspace{-0.2cm}
 	\caption{
 	Model architecture of LLM-based multi-talker ASR
 	}
 	\label{fig:llm-sot}
 	\vspace{-0.2cm}
\end{figure}

\vspace{-0.3cm}
\subsection{LLM-Based SOT for Multi-Talker ASR}
{\color{black}In previous works, attention-based encoder-decoder (AED) architectures have been employed to implement SOT-based multi-talker ASR. Considering that SOT-style transcription involves concatenating potentially related utterances from multiple speakers, the model requires strong long-context awareness and the ability to model across utterances. Unlike AED architectures that use cross attention to obtain recognition sequences, LLM architectures directly utilize their powerful decoders, which have undergone extensive pre-training, to generate text. Therefore, LLM-based models are likely better suited for this complex and challenging task.}
Given these considerations, we propose an LLM-based model to further overcome the performance bottlenecks of SOT-based multi-talker ASR.

As shown in Fig.~\ref{fig:llm-sot}, the architecture for LLM-based multi-talker ASR mainly consists of a speech encoder, a projector, and an LLM. For each sample, given the overlapped speech signal $S_\text{olp}$ and the corresponding SOT-style multi-talker transcription $T_\text{multi}$, a speech encoder is first used to convert the overlapped speech signal into a speech representation, which can be represented as:
\begin{align}
H^s=\text{Encoder}(S_\text{olp})
\end{align}
$H^s \in \mathbb{R}^{f^s \times l^s}$ is the speech representation, where $f^s$ and $l^s$ denote the feature dimension and the length, respectively. $H^s$ can be very long, making it difficult for the LLM to process and increasing the computational burden. Therefore, we stack every $n$ consecutive frames in the feature dimension to downsample the representation, denoted as:
\begin{align}
\bar{H}^s = \text{Downsampler}(H^s)
\end{align}
where $\bar{H}^s \in \mathbb{R}^{(f^s \cdot n) \times l^{\bar{s}}}$ is the output after downsampling. The length of $\bar{H}^s$ is $l^{\bar{s}}$, which is more suitable for the LLM. The dimension of the speech representation is expanded by a factor of $n$. Then, a projector is introduced to convert the speech representation into a speech embedding that resides in the same domain as the text embedding and has the same dimension as the hidden size of the LLM, denoted as:
\begin{align}
E^s=\text{Projector}(\bar{H}^s)
\end{align}
We tokenize the SOT-style multi-talker transcription and obtain the text embedding $E^t$, denoted as:
\begin{align}
E^t = \text{Embedding}(\text{Tokenizer}(T_\text{multi}))
\end{align}
Finally, during the training stage, the speech embedding and text embedding are concatenated together as the input to the LLM. The output of the LLM is the predicted SOT-style multi-talker transcription $\hat{T}_\text{multi}$, denoted as:
\begin{align}
\hat{T}_\text{multi} = \text{LLM}(\text{Concat}(E^s, E^t))
\end{align}
Cross-Entropy (CE) is used as the loss function:
\begin{align}
\mathcal{L} = \text{CE}(\hat{T}_\text{multi}, T_\text{multi})
\end{align}
Since the begin ($\langle\text{bos}\rangle$) and end token ($\langle\text{eos}\rangle$) are introduced during training, the speech embedding is used as the input to the LLM during the inference stage, allowing the multi-talker transcription to be predicted via auto-regressive inference.

\section{Experiments}
\vspace{-0.2cm}
We first conducted experiments on the modified simulated dataset LibriMix~\cite{cosentino2020librimix}, where each utterance contains only 2 speakers with a time delay between them. Then, we evaluated our model on the real-world meeting scenario dataset AMI~\cite{CarlettaABFGHKKKKLLLMPRW05}, where each meeting in the evaluation set contains up to 4 speakers.

\subsection{Experiment with LibriMix}
\subsubsection{Dateset and evaluation metric}
We used LibriMix\footnote{\url{https://github.com/espnet/espnet/tree/master/egs2/librimix/sot_asr1}\label{espnet}} modified by ESPnet~\cite{WatanabeHKHNUSH18} for preliminary experiments. LibriMix is a simulated dataset obtained by mixing single-speaker speech from LibriSpeech~\cite{PanayotovCPK15} with noise from WHAM!~\cite{WichernAFZMCMR19,MaciejewskiWMR20}. The official LibriMix is used for the source separation task, where the simulation process typically assumes fully-overlapped speech, meaning that speech from different speakers starts at the same time. To make it suitable for the multi-talker ASR task, the original simulation process is modified in the ESPnet pipeline\footref{espnet} to introduce a random delay ranging from 1 to 1.5 seconds for the mixed speech. The final generated simulated data contains approximately 830 hours of speed-perturbed training set, 8.2 hours of development set, and 7.6 hours of test set, with two speakers in all utterances.

In the LibriMix experiment, to compare with the results from ESPnet, we used word error rate (WER) as the evaluation metric. This metric is directly calculated between the predicted and reference SOT-style multi-talker transcriptions.

\subsubsection{Model configuration}
We utilized WavLM\footnote{\url{https://huggingface.co/microsoft/wavlm-large}}~\cite{ChenWCWLCLKYXWZ22} as the speech encoder because both the Base+ and Large versions of WavLM leverage a substantial amount of overlapped speech data for self-supervised pre-training, making them suitable for the multi-talker ASR task. The LLM module chosen was Vicuna-7B\footnote{\url{https://huggingface.co/lmsys/vicuna-7b-v1.5}}~\cite{chiang2023vicuna}, a chat model fine-tuned from the pre-trained LLaMA~\cite{abs-2302-13971,abs-2307-09288} on conversational data collected from ShareGPT users. The downsampling rate $n$ was set to 10, resulting in speech embedding with frames of 200 ms length. Two linear layers acted as projectors with ReLU activation in between, and the hidden size was set to 4096. {\color{black}We used Vicuna tokenizer in all systems.}

\vspace{-0.2cm}
\subsubsection{Training strategy and detail}
In previous works on LLM-based ASR, different training strategies were employed. In~\cite{abs-2307-11795}, the speech encoder, projector, and LoRA adaptor were trained together. In~\cite{abs-2402-08846}, the LoRA was not introduced, the speech encoder was frozen, and only the projector was trained. In~\cite{abs-2405-02132}, the three modules were unfrozen in three stages, following the order of projector $\rightarrow$ speech encoder $\rightarrow$ LoRA. In the LibriMix experiment, we adopted a multi-stage training strategy similar to that in \cite{abs-2405-02132}. The benefit of this multi-stage training is that it enhances the model's capacity to align auditory and textual information. A slight difference in our approach is that when using the WavLM model fine-tuned with LibriMix, the training process requires freezing the speech encoder.

We used 8 NVIDIA V100 32GB GPUs for training, with a batch size of 2 samples per GPU and a gradient accumulation of 4. The DeepSpeed strategy~\cite{RasleyRRH20} was used for distributed training. We employed the AdamW optimizer~\cite{LoshchilovH19} with a learning rate of 0.0001, betas of (0.9, 0.999), epsilon of 1e-08, and weight decay of 1e-6. A linear warmup scheduler was used, with 2000 warmup steps and a maximum of 100,000 training steps, but training was stopped early if the validation loss did not decrease. We applied this training configuration in each training stage. When training the LLM, we only performed LoRA fine-tuning with alpha = 16 and rank = 16. In all experiments, greedy search was used for decoding.

\vspace{-0.2cm}
\subsubsection{Experimental results}
Table~\ref{tab:lm_overall} shows our results comparing various approaches on LibriMix. Sys. \{1-3\} are the results from ESPnet. Among these, using a conformer as the encoder and the WavLM Large model as upstream achieves better results because the WavLM model has been self-supervised pre-trained on large-scale overlapped speech, making it more suitable for multi-talker scenarios. Sys. \{4-5\} in Table~\ref{tab:lm_overall} are the results of fine-tuning the WavLM model using AED approach. The performance of the WavLM Large model is significantly better than that in ESPnet, since the WavLM in the latter is frozen. Sys. \{6-8\} in Table~\ref{tab:lm_overall} are the results of the LLM-based approach proposed in this work. 
{\color{black}When using WavLM Base+ as the speech encoder, the LLM-based method (Sys. 6, Tab.~\ref{tab:lm_overall}) outperforms the AED-based method (Sys. 4, Tab.~\ref{tab:lm_overall}). However, when WavLM Large is used as the encoder, the AED-based method shows a significant performance boost (Sys. 5, Tab.~\ref{tab:lm_overall}), even surpassing the LLM-based method (Sys. 7, Tab.~\ref{tab:lm_overall}), which indicates that AED-based systems are more dependent on encoder performance. Initializing the LLM-based system with the speech encoder fine-tuned on LibriMix using AED method results in the best performance (Sys. 8, Tab.~\ref{tab:lm_overall}). Therefore, in the performance on LibriMix test set, the advantage of the LLM-based system over the AED-based system is not very pronounced (9.0\% WER in Sys. 8 \textit{vs.} 9.2\% WER in Sys. 5). This is similar to conclusions drawn from single-speaker ASR studies~\cite{abs-2307-11795,abs-2402-08846}, as LibriMix is simulated data and contains only two speakers per utterance, making it less challenging compared to real conversational scenarios.}

\begin{table}[t!]
\renewcommand{\thetable}{1}
\centering
\caption{Overall performance comparison of various approaches on LibriMix. {\color{black}Sys. \{1-3\} are the experimental results from ESPnet\footref{espnet}, Sys. \{4-5\} are the results of AED-based models, and Sys. \{6-8\} are the results of the LLM-based models.}}
\setlength{\tabcolsep}{2mm}
\resizebox{\columnwidth}{!}{
\begin{tabular}{c|c|l|cc}
\toprule
\hline
\multirow{2}{*}{Sys.} & \multirow{2}{*}{type} & \multicolumn{1}{c|}{\multirow{2}{*}{Speech Encoder}} & \multicolumn{2}{c}{WER~(\%)  $\downarrow$} \\
 & & & dev & test \\ \hline
1 & \multirow{3}{*}{\shortstack{ESPnet\footref{espnet} \\ Baseline}} & Whisper small & 26.0 & 25.0 \\
2 & & Conformer & 24.7 & 23.3 \\
3 & & \quad + WavLM Large upstream & 19.4 & 17.1 \\ \hline
4 & \multirow{2}{*}{AED} & WavLM Base+ & 18.9 & 17.7 \\
5 & & WavLM Large & \color{black}10.6 & \color{black}9.2 \\ \hline
6 & \multirow{3}{*}{LLM} & WavLM Base+ & 17.6 & 15.9 \\
7 & & WavLM Large & 11.4 & 10.2 \\
8 & & \quad + LibriMix Fine-tuning & \textbf{10.3} & \textbf{9.0} \\ \hline
\bottomrule
\end{tabular}
}
\label{tab:lm_overall}
\end{table}


\begin{table}[t]
\renewcommand{\thetable}{2}
\centering
\caption{\color{black}Performance comparison with and without LoRA fine-tuning in the case of different speech encoders.}
\setlength{\tabcolsep}{1.5mm}
\resizebox{\columnwidth}{!}{\begin{tabular}{c|l|c|cc}
\toprule
\hline
\multicolumn{1}{c|}{\multirow{2}{*}{Sys.}} & \multicolumn{1}{c|}{\multirow{2}{*}{Speech Encoder}} & \multirow{2}{*}{LoRA} & \multicolumn{2}{c}{WER~(\%) $\downarrow$} \\
 & & & dev & test \\ \hline
- & \multirow{2}{*}{WavLM Base+} & \xmark & 19.4 & 17.3 \\
Tab.~\ref{tab:lm_overall}, Sys. 6 & & \cmark & 17.6 & 15.9 \\ \hline
- & \multirow{2}{*}{WavLM Large} & \xmark & 12.6 & 11.3 \\
Tab.~\ref{tab:lm_overall}, Sys. 7 & & \cmark & 11.4 & 10.2 \\ \hline
- & \multirow{2}{*}{\quad + LibriMix Fine-tuning} & \xmark & 10.8 & 9.5 \\
Tab.~\ref{tab:lm_overall}, Sys. 8 & & \cmark & 10.3 & 9.0 \\ \hline
\bottomrule
\end{tabular}}
\label{tab:compare_lora}
\end{table}

\begin{table}[t!]
\renewcommand{\thetable}{3}
\centering
\caption{\color{black}Performance comparison of freezing and jointly training the speech encoder with and without fine-tuning on LibriMix using AED method.}
\setlength{\tabcolsep}{3.5mm}
\resizebox{\columnwidth}{!}{\begin{tabular}{c|c|c|cc}
\toprule
\hline
\multicolumn{1}{c|}{\multirow{2}{*}{Sys.}} & \multirow{2}{*}{\shortstack{LibriMix\\Fine-tuning}} & \multirow{2}{*}{\shortstack{Freeze\\Encoder}} & \multicolumn{2}{c}{WER~(\%) $\downarrow$} \\
 & & & dev & test \\ \hline
Tab.~\ref{tab:lm_overall}, Sys. 7 & \multirow{2}{*}{\xmark} & \xmark & 11.4 & 10.2 \\
- & & \cmark & 47.8 & 46.7 \\ \hline
- & \multirow{2}{*}{\cmark} & \xmark & 11.4 & 10.1 \\
Tab.~\ref{tab:lm_overall}, Sys. 8 & & \cmark & 10.3 & 9.0 \\ \hline
\bottomrule
\end{tabular}}
\label{tab:encoder_freeze}
\end{table}

\begin{table}[t!]
\renewcommand{\thetable}{4}
\centering
\caption{Performance comparison of single-stage training and multi-stage training strategy. {\color{black}Multi-stage training refers to sequentially unfreezing and jointly training in the order of projector $\rightarrow$ speech encoder $\rightarrow$ LoRA. When the ``Freeze Encoder'' option in the table is set to True, the second stage is skipped. Single-stage training refers to jointly training all these modules from the beginning.}}
\setlength{\tabcolsep}{3mm}
\resizebox{\columnwidth}{!}{\begin{tabular}{c|c|c|cc}
\toprule
\hline
\multicolumn{1}{c|}{\multirow{2}{*}{Sys.}} & \multirow{2}{*}{\shortstack{Freeze\\Encoder}} & \multirow{2}{*}{\shortstack{Training\\Strategy}} & \multicolumn{2}{c}{WER~(\%) $\downarrow$} \\
 & & & dev & test \\ \hline
- & \multirow{2}{*}{\xmark} & single-stage & 11.7 & 10.4 \\
- & & multi-stage & 11.4 & 10.1 \\ \hline
- & \multirow{2}{*}{\cmark} & single-stage & 10.5 & 9.2 \\
Tab.~\ref{tab:lm_overall}, Sys. 8 & & multi-stage & 10.3 & 9.0 \\ \hline
\bottomrule
\end{tabular}}
\label{tab:train_strategy}
\end{table}

Table~\ref{tab:compare_lora} shows the performance comparison of different speech encoders with and without LoRA fine-tuning. Similar to the conclusions in \cite{abs-2307-11795} and \cite{abs-2405-02132}, introducing LoRA fine-tuning into the LLM consistently improves performance regardless of the speech encoder used. This indicates that LoRA fine-tuning can adapt the LLM output to the style of SOT-based multi-talker transcription. In \cite{abs-2402-08846}, promising performance can be achieved even without introducing the LoRA adaptor, possibly because the transcription style of the single-talker Librispeech used in \cite{abs-2402-08846} is similar to the output of the original LLM.

\begin{figure}[tp]
 	\centering
 	\includegraphics[width=1\linewidth]{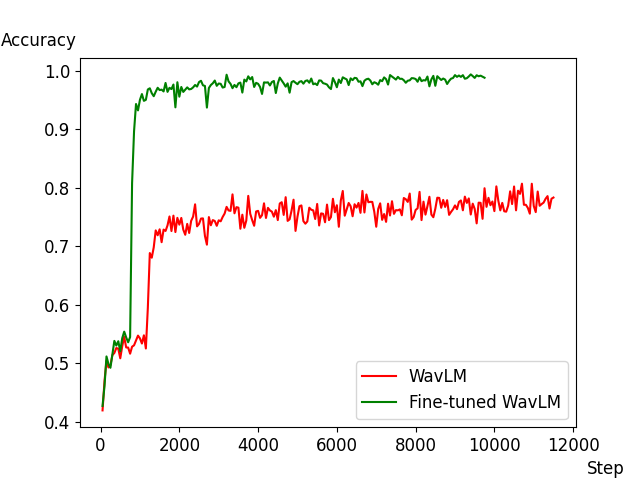}
 	\caption{
 	 Training accuracy of the next token prediction with the training steps in the first training stage, where only the Projector is involved in training. Different colored curves represent whether the speech encoder has been fine-tuned by LibriMix.
 	}
 	\label{fig:acc_compare}
\end{figure}

Table~\ref{tab:encoder_freeze} presents the impact of freezing the speech encoder during training. When the initialized speech encoder is not fine-tuned on LibriMix using the AED method, freezing the encoder results in poor performance because the encoder has not adapted to the LibriMix dataset. However, when using the encoder fine-tuned with LibriMix (Sys. 5, Tab.~\ref{tab:lm_overall}), freezing the encoder during training results in better performance. This is likely because the fine-tuned encoder already has excellent representation extraction capabilities on LibriMix and does not require further adjustment.

Fig.~\ref{fig:acc_compare} shows the comparison of training curves in the first training stage, where only the projector module is trained, using either a fine-tuned encoder or a non-fine-tuned encoder. When using the fine-tuned encoder, the model quickly converges to a very high accuracy. In contrast, using the original WavLM model results in slower and less complete convergence. This indicates that if we have a high-quality encoder, simply aligning the modality of speech representations with the LLM can directly achieve a relatively good performance. Conversely, for an unadapted encoder, merely training the projector to perform alignment is insufficient, which is similar to the conclusion in Table~\ref{tab:encoder_freeze}.

Table~\ref{tab:train_strategy} presents the comparison between single-stage and multi-stage training strategies. The results show that, regardless of whether the speech encoder is frozen, multi-stage training outperforms single-stage training. This indicates that multi-stage training helps the model better align auditory and textual information.

\subsection{Experiment with AMI}
\subsubsection{Experimental settings}
To evaluate the LLM-based multi-talker ASR approach in a more realistic setting, we conducted experiments on real-world corpus AMI. The AMI meeting corpus includes approximately 95 hours of real-world meeting recordings, with the training, validation, and evaluation sets comprising 76.9, 8.9, and 8.7 hours, respectively. Each meeting involves 3 to 5 participants. The audio in the AMI corpus was recorded using an 8-channel microphone array, known as multiple distant microphones (MDM). Typically, the first channel is used for monaural ASR evaluation, referred to as the single distant microphone (SDM) setting. Additionally, the AMI corpus includes near-field single-speaker audio recorded by independent headset microphones (IHM) worn by each participant.

In this work, we conducted experiments using the SDM setting. However, in the original SDM, the audio is segmented by oracle timestamps into utterances containing only a single speaker. To evaluate SOT-based multi-talker ASR, we followed the approach in~\cite{KandaYWGWMCY21} to use \textit{utterance group}-based evaluation. An \textit{utterance group} is defined as a set of utterances connected by speaker overlap regions. Correspondingly, SOT-style transcriptions are generated in the order of the emission time of each speaker.

In addition to using simple WER for evaluation, we also introduced the concatenated minimum-permutation word error rate (cpWER)~\cite{abs-2004-09249} for comparison with previous work~\cite{KandaYWGWMCY21,LiQ0KWYQ023}. In each \textit{utterance group}, {\color{black}as shown in Fig.~\ref{fig:sot}}, the transcriptions of the same speaker are concatenated, and the minimum WER across all possible speaker permutations is taken as the cpWER.

\begin{figure}[tp]
 	\centering
 	\includegraphics[width=0.9\linewidth]{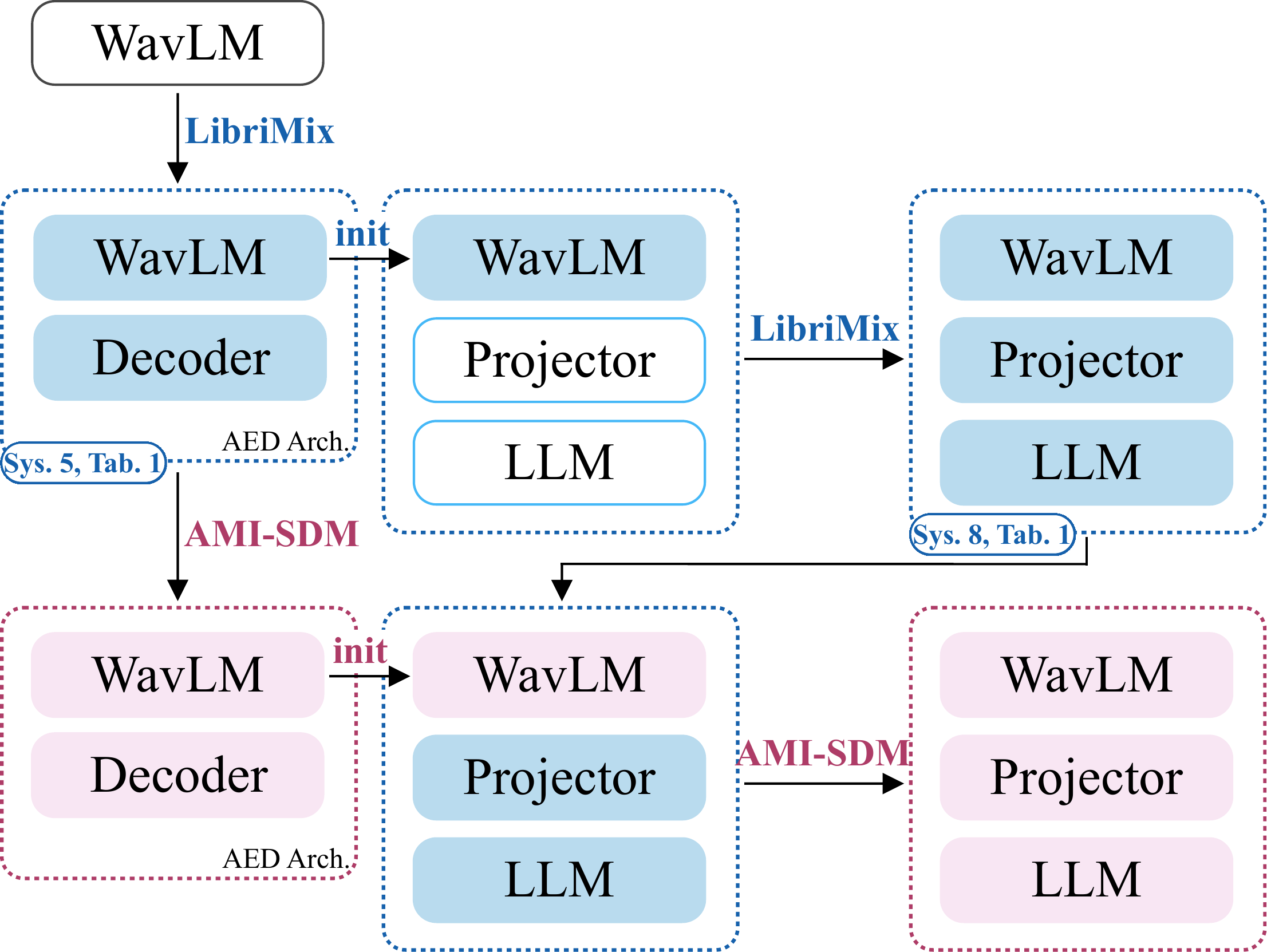}
        \caption{An illustration of the training process of the proposed LLM-based multi-talker ASR system on the LibriMix (blue background) and AMI-SDM (pink background).}
 	\label{fig:train_flow}
\end{figure}

For the training details, as shown in Fig~\ref{fig:train_flow}, we first fine-tuned the WavLM AED model, which was pre-trained on LibriMix (Sys. 5, Tab.~\ref{tab:lm_overall}), using the AMI-SDM \textit{utterance group} segments. Subsequently, we integrated this fine-tuned WavLM encoder into the best-performing system from the LibriMix experiment (Sys. 8, Tab.~\ref{tab:lm_overall}) and further fine-tuned it on the AMI-SDM \textit{utterance group} segments. The training strategy and configuration remained consistent with those employed in the LibriMix experiment.

\begin{table*}[t!]
\centering
\caption{Overall performance comparison of various approaches on AMI-SDM evaluation set. {\color{black}Sys. \{1-3\} are previous works that use large-scale supervised data for pre-training. Sys. \{4-6\} display the results of models pre-trained with only 0.83k hours of LibriMix and then fine-tuned on AMI, where Sys. 4 uses the AED-based architecture, and Sys. \{5-6\} use the LLM-based architecture. The WER (\%) and cpWER (\%) metrics are reported for the \textit{utterance groups} with different numbers of speakers, as well as overall (average) results.}}
\setlength{\tabcolsep}{2mm}
\resizebox{\textwidth}{!}{
\begin{tabular}{c|l|c|c|ccccc|ccccc}
\toprule
\hline
\multirow{2}{*}{Sys.} & \multicolumn{1}{c|}{\multirow{2}{*}{Architecture}} & \multirow{2}{*}{\shortstack{Supervised\\Pre-training  data}} & \multirow{2}{*}{\shortstack{Fine-tuning\\data}} & \multicolumn{5}{c|}{WER (w.r.t. \# of talkers) (\%) $\downarrow$} & \multicolumn{5}{c}{cpWER (w.r.t. \# of talkers) (\%) $\downarrow$} \\
 & & & & avg. & 1 & 2 & 3 & 4 & avg. & 1 & 2 & 3 & 4 \\ \hline
 1 & Conformer AED~\cite{KandaYWGWMCY21} & 900k hrs & \multirow{6}{*}{AMI} & - & - & - & - & - & 21.2 & 14.7 & 19.6 & \textbf{25.7} & \textbf{35.5} \\ \cline{1-3} \cline{5-14}
 2 & Whisper medium~\cite{LiQ0KWYQ023} & \multirow{2}{*}{680k hrs} & & - & - & - & - & - & 23.6 & 12.8 & 21.8 & 32.5 & 45.9 \\
 3 & Whisper large~\cite{LiQ0KWYQ023} &  &  & - & - & - & - & - & 21.4 & 12.0 & 20.0 & 29.3 & 40.6 \\ \cline{1-3} \cline{5-14}
 4 & WavLM Large AED & \multirow{3}{*}{\textbf{0.83k hrs}} & & 30.5 & 16.7 & 26.2 & 45.8 & 54.8 & 24.1 & 10.8 & 20.4 & 37.9 & 48.6 \\
 5 & WavLM Large LLM &  &  & 27.6 & 14.9 & 25.3 & 38.4 & 52.6 & 21.0 & 9.3 & 18.8 & 31.1 & 44.1 \\
 6 & $\quad$ + beam search (beam=4) &  &  & \textbf{26.8} & \textbf{14.8} & \textbf{24.4} & \textbf{37.5} & \textbf{49.4} & \textbf{20.4} & \textbf{9.3} & \textbf{18.1} & 30.3 & 42.2 \\ \hline
\bottomrule
\end{tabular}}
\vspace{-2mm}
\label{tab:ami_overall_comparison}
\end{table*}

\begin{table}[t!]
\centering
\caption{\color{black}Speaker counting accuracy (\%) for each \textit{utterance group} of AMI-SDM evaluation set. The number of talkers can be estimated by counting the segments obtained by separating SOT-style transcriptions with the speaker change symbol ``\$''. SOT follows speaker-wise FIFO, as shown in Fig.~\ref{fig:sot}.}
\resizebox{\columnwidth}{!}{
\begin{tabular}{c|c|cccccc}
\toprule
\hline
\multirow{2}{*}{Sys.} & \multirow{2}{*}{\shortstack{Actual \#\\of talkers}} & \multicolumn{6}{c}{Estimated \# of talkers} \\
 & & 0 & 1 & 2 & 3 & 4 & $\geq$ 5 \\ \hline
\multirow{4}{*}{Tab.~\ref{tab:ami_overall_comparison}, Sys. 1} & 1 & 0.2 & \textbf{97.2} & 2.5 & 0.1 & 0.0 & 0.0 \\
 & 2 & 0.0 & 13.7 & \textbf{80.5} & 5.9 & 0.0 & 0.0 \\
 & 3 & 0.0 & 2.4 & 32.6 & \textbf{60.2} & 4.8 & 0.0 \\
 & 4 & 0.0 & 0.0 & 9.9 & 51.2 & \textbf{38.9} & 0.0 \\ \hline
\multirow{4}{*}{Tab.~\ref{tab:ami_overall_comparison}, Sys. 4} & 1 & 0.0 & \textbf{92.1} & 7.6 & 0.3 & 0.0 & 0.0 \\
 & 2 & 0.0 & 10.0 & \textbf{73.0} & 16.4 & 0.6 & 0.0 \\
 & 3 & 0.0 & 0.8 & 30.2 & \textbf{58.0} & 10.1 & 0.9 \\
 & 4 & 0.0 & 0.0 & 5.5 & 52.0 & \textbf{33.5} & 9.0 \\ \hline
\multirow{4}{*}{Tab.~\ref{tab:ami_overall_comparison}, Sys. 5} & 1 & 0.0 & \textbf{96.7} & 3.2 & 0.1 & 0.0 & 0.0 \\
 & 2 & 0.0 & 11.7 & \textbf{76.9} & 11.0 & 0.4 & 0.0 \\
 & 3 & 0.0 & 1.3 & 39.3 & \textbf{48.4} & 10.8 & 0.2 \\
 & 4 & 0.0 & 0.0 & 10.5 & 53.0 & \textbf{35.0} & 1.5 \\ \hline
\bottomrule
\end{tabular}}
\vspace{-1mm}
\label{tab:ami_spk_comparison}
\end{table}

\subsubsection{Experimental results}
The overall experimental results on the AMI-SDM evaluation set are presented in Table~\ref{tab:ami_overall_comparison}. Sys. \{1-3\} are from previous work, all relying on large-scale supervised data for pre-training. As shown by the experimental results, in terms of the average cpWER metric, the LLM-based approach (Sys. 5, Tab.~\ref{tab:ami_overall_comparison}) not only outperforms the AED-based method using the same amount of data (Sys. 4, Tab.~\ref{tab:ami_overall_comparison}) but also remarkably surpasses the models in Sys. \{1-3\} that were trained with large-scale supervised data. It is worth mentioning that Sys. 1 in Table~\ref{tab:ami_overall_comparison} was trained using 900k hours of supervised data, which is 1000 times more than what we used.
{\color{black}This demonstrates that for SOT-based multi-talker ASR task, having a robust, large-scale pre-trained decoder is more important, as it provides strong capabilities in long-context awareness and cross-utterance modeling. This is precisely the advantage of LLM-based architectures over traditional AED-based systems in a such complex scenarios involving multi-talker conversations. 
\textcolor{black}{Additionally, the performance advancement of the LLM-based model over the AED-based method is further highlighted on the AMI evaluation set with an absolute WER reduction of 2.9\% and cpWER reduction of 3.1\% (Sys. 5 \textit{vs.} 4, Tab. \ref{tab:ami_overall_comparison}, column ``avg.'') comparing with the comparable systems evaluated on the LibriMix test set (Sys. 8 \textit{vs.} 5, Tab. \ref{tab:lm_overall}).}
This indicates that the more realistic and complex the scenario, the greater the advantage of the LLM-based method, confirming our conjecture.} Using beam search for decoding yields even better results (Sys. 6, Tab.~\ref{tab:ami_overall_comparison}).

Comparing the results across \textit{utterance groups} with different numbers of speakers, we find that the LLM-based method performs worse than Sys. 1 and Sys. 3 in groups with 3 and 4 speakers. This may be due to the limited supervised training data used in the LLM-based method, especially since the LibriMix dataset used for pre-training only contains two-speaker utterances, and the AMI training set has relatively few \textit{utterance groups} with more than 2 speakers. 
{\color{black}In Sys. 2, the speech encoder in the Whisper medium model~\cite{RadfordKXBMS23} has a parameter amount very close to that of WavLM Large, and the decoder of Whisper is also large. However, the LLM-based method consistently outperforms in \textit{utterance groups} containing any number of speakers. This once again highlights the superiority of the LLM-based architecture, which leverages a powerful pre-trained decoder, over the AED-based architecture, where the decoder has not undergone specialized pre-training, in recognizing SOT-style long transcriptions with related content from multiple speakers.}

We calculated the speaker counting accuracy and presented it in Table~\ref{tab:ami_spk_comparison}. From the results, it can be observed that the LLM-based method (Tab.~\ref{tab:ami_overall_comparison}, Sys. 5) is less accurate in estimating the number of speakers compared to the AED model trained with large-scale supervised data (Tab.~\ref{tab:ami_overall_comparison}, Sys. 1). Additionally, in the cases of 3 and 4 speakers, it also shows no significant advantage over the AED model using the same amount of data (Tab.~\ref{tab:ami_overall_comparison}, Sys. 4). Despite the lower accuracy in speaker counting, the LLM-based method achieves the best performance in the cpWER metric, indicating that it has a very high accuracy in recognizing the content of transcriptions in complex scenarios involving multi-talker conversations with noise and reverberation.

\section{Conclusions}
\vspace{-0.2cm}
In this paper, we pioneer an LLM-based multi-talker ASR approach. In the evaluation, the proposed method achieves state-of-the-art results on both the simulated data LibriMix and the real-world data AMI, even outperforming existing methods trained with 1000 times more supervised data on the AMI-SDM evaluation set. {\color{black}The experimental results demonstrate that LLM-based architectures, which emphasize decoder performance and possess strong capabilities in understanding long contexts and modeling across utterances, outperform AED-based structures that focus more on encoder performance in SOT-based multi-talker ASR task. The LLM-based method has a much larger advantage on real data AMI than on simulated data LibriMix,} which further highlights the potential of LLM-based models in handling speech processing tasks in complex and challenging scenarios.



\newpage


\end{document}